\documentclass[11pt,twoside]{article}

\usepackage{asp2006}
\usepackage{graphicx}
\usepackage{amssymb}

\begin{document}

\title{MASS SEGREGATION IN THE GALACTIC CENTRE} 

\author{Clovis Hopman and Ann-Marie Madigan}

\affil{Leiden Observatory, Leiden University, P.O. Box 9513, NL-2300 RA Leiden}
 
\begin{abstract} 
Two-body energy exchange between stars orbiting massive black holes (MBHs) leads to the formation of a power-law density distribution $n(r)\propto r^{-\alpha}$ that diverges towards the MBH. For a single mass population, $\alpha=7/4$ and the flow of stars is much less than $N(<r)/t_r$ (enclosed number of stars per relaxation time). This ``zero-flow" solution is maintained for a multi-mass system for moderate mass ratios or systems where there are many heavy stars, and slopes of $3/2<\alpha<2$ are reached, with steeper slopes for the more massive stars. If the heavy stars are rare and massive however, the zero-flow limit breaks down and much steeper distributions are obtained. 

We discuss the physics driving mass-segregation with the use of Fokker-Planck calculations, and show that steady state is reached in $0.2-0.3 ~ t_r$. Since the relaxation time in the Galactic centre (GC) is $t_r \sim 2-3 \times 10^{10} {\rm yr}$, a cusp should form in less than a Hubble time. The absence of a visible cusp of old stars in the GC poses a challenge to these models, suggesting that processes other than two-body relaxation have played a role. We discuss astrophysical processes within the GC that depend crucially on the details of the stellar cusp.
\end{abstract}

\section{Introduction}

\subsection{Historical overview of the theory}

The formation of a stellar cusp of stars near massive black holes (MBHs) was first discussed by \citet{Pee72}, who argued that a Maxwellian distribution is not feasible because stars would be destroyed at unphysical high rates close to the MBH. \citet{Bah76} performed a more detailed calculation and showed that indeed a cusp forms, but with a shallower slope than found by Peebles. The Bahcall \& Wolf profile of $n(r)\propto r^{-7/4}$, where $n(r)$ is the density at radius $r$, has since been confirmed in many theoretical papers with different methods, including Fokker-Planck \citep[e.g.][]{Coh78, Mur91}, Monte Carlo \citep[e.g.][]{Sha78, Fre02} and $N$-body methods \citep{Pre04, Bau04a}, and can be understood (with hindsight) by simple dimensional arguments \citep[e.g.][]{Bin08, Sar06}.

In a second paper, \citet{Bah77} considered the evolution of stellar systems with several masses (denoted with $M$) and showed that again the resulting steady state distribution functions $f_{M}(E) \propto E^{p_M}$ are approximate power laws, with the more massive stars having steeper slopes. They found that the stellar current into the MBH, $Q_{M} (E )$, is very small (``zero-flow solution") which leads to a specific relation between the stellar mass and the logarithmic slope of the DF\footnote{Throughout this paper the subscripts $L, H$ refer to either the light or the heavy component of a two-mass system.}, $p_M = M_L/4M_H$. In the Keplerian limit near the MBH, these DFs correspond to power-law density cusps, $n(r)\propto r^{-\alpha_M}$ with $\alpha_M = 3/2 + p_M$. In their analysis, they considered quite moderate number and mass-ratios, and concluded that the stars with the lowest masses have slopes $\alpha_L \gtrsim 3/2$, whereas the stars with the highest masses have slopes $\alpha_H \lesssim 2 $. In contrast to their work on single mass-systems, for several decades surprisingly few studies pursued the dynamics of multi-mass systems. 

\citet{Mur91} performed a Fokker-Planck study with a mass-spectrum, in addition to other physical processes such as stellar collisions, tidal disruptions and stellar evolution. More recently, motivated by observations of the stellar population surrounding the MBH in Galactic centre (GC), there have been several studies of mass-segregation near MBHs. The first applications to the GC were by \citet{Fre02, Fre03, Fre06}, who used H\'enon-type Monte Carlo simulations. These simulations broadly confirmed the results by \citet{Bah77}, and showed that cusps should indeed form near the MBH in our own GC.

The planning of the {\it Laser Interferometer Space Antenna} (LISA) formed another motivation for the study of mass-segregation near MBHs \citep[e.g.,][]{Sig97b}. 
The reason for this is as follows. When compact stellar remnants spiral into MBHs of $\sim10^6 {M_{\odot}}$, they emit gravitational waves with frequencies and power such that LISA should be able to observe them out to redshifts $z\gtrsim1$. It was shown by \citet{Hop05} that stars can only spiral in successfully if they start very close ($\lesssim0.01 {\rm pc}$) to the MBH, for otherwise they are scattered away from their inspiral orbit. The implication is that inspiral rates are very sensitive to the details of the stellar distribution very close to the MBH, which is in turn dependent on the amount of mass segregation in the cusp. As a result, inspiral rates of stellar black holes (BHs) may be higher than those of white dwarfs (WDs) \citep{Hop06b}: even though for typical initial mass functions there are many more WDs than BHs in an evolved stellar population, mass-segregation drives the BHs to such orbits where they can spiral in, such that within $\sim0.01{\rm pc}$ there may in fact be more BHs than WDs. In addition to inspiral stars, LISA may also observe gravitational wave bursts from stars during a single fly-by in our {\it own} GC \citep{Rub06}. These stars are usually on very short period ($\sim1{\rm yr}$) orbits, and mass-segregation makes it again more likely to observe bursts from BHs than from other stars \citep{Hop07}.

The interest in mass segregation near MBHs, driven by observation, has also led to new theoretical developments. \citet{Ale09} extended the analysis by \citet{Bah77} to a much larger parameter space, and showed that the study by \citet{Bah77} happened to be at a corner of this space where the behaviour is very stable, but that for systems where the massive stars are less numerous the evolution is very different; see \S\ref{s:duo} and Figure (\ref{f:Delta}). More specifically, they found that the most massive stars can have power law slopes that are steeper than $\alpha=2$, which was ruled out by \citet{Bah77}. This kind of ``strong mass segregation" was confirmed in $N$-body simulations by \citet{Pre10}. An analytical study of mass segregation with {\it continuous} mass functions was performed by \citet{Kes09}, who showed that, in the maximally steep limit, heavy stars can develop a cusp as steep as $n(r) \propto r^{-3}$.

In this work we revisit the evolution of a cusp from a duo-mass stellar population which undergoes mass segregation \citep{Ale09}, determine the time scale for arriving at a steady state distribution, and discuss the results in terms of the GC. 

\section{Duo-mass systems and strong mass-segregation}\label{s:duo}
\citet{Ale09} use Fokker-Planck simulations to find the the approximate steady state distribution functions (DFs) of a non-evolving spherically-distributed duo-mass stellar population around a fixed MBH. Mass is measured in units of the mass of a reference star $M_{\star}$, specific energy in units of its velocity dispersion $\epsilon_{\star}\!=\!\sigma_{\star}^{2}$, and time in units of its two-body relaxation time at the radius of influence, 
\begin{equation}\label{e:tstar}
t_{\star}=\frac{3(2\pi\sigma_{\star}^{2})^{3/2}}{32\pi^{2}G^{2}M_{\star}^{2}n_{\star}\ln\Lambda}\,,
\end{equation}
where the Coulomb term is estimated as $\Lambda\!=\!M_{\bullet}/M_{\star}$. Phase
space density is expressed in units of $f_{\star}\!=\! n_{\star}/(2\pi\sigma_{\star}^{2})^{3/2}$
and distance in units of the MBH radius of influence $r_{\star}\!=\! GM_{\bullet}/\sigma_{\star}^{2}$.
In these units, the dimensionless specific orbital energy is defined
as $x\!=\!\epsilon/\sigma_{\star}^{2}\!=\!r_{\star}/(2a)$ ($a$ is the
semi-major axis), the dimensionless time is defined as $\tau\!=\! t/t_{\star}$,
and the dimensionless DF of each stellar mass group as $g_{M}\!=\! f_{M}/f_{\star}$. The evolution of the dimensionless DF, $g_{M}$, is given by the time and energy dependent particle conservation equation that describes the two-body diffusion of stars in energy from a fixed unbound reservoir into the MBH sink,
\begin{equation}
\frac{\partial}{\partial\tau}g_{M}(x,\tau)=-x^{5/2}\frac{\partial}{\partial x}Q_{M}(x,\tau)\,.\label{e:FPeq}\end{equation}
$Q_{M}$ is the energy flow integral which expresses the change in energy due to two-body scattering \citep{Bah76,Bah77},
\begin{eqnarray}
Q_{M}(x) & = & \sum_{M'}MM'\int_{-\infty}^{x_{D}}\frac{\mathsf{d}x'}{\max\left(x,x'\right)^{3/2}}\times\nonumber \\
 &  & \left[g_{M}(x)\frac{\partial g_{M'}(x')}{\partial x'}\!-\!\frac{M'}{M}g_{M'}(x')\frac{\partial g_{M}(x)}{\partial x}\right]\,.\label{e:Qm}\end{eqnarray}
Equation (\ref{e:FPeq}) is integrated in time from an arbitrary initial
DF until steady state is achieved, subject to the boundary conditions
that the DF falls to zero at some very high energy $x_{D}$ where
the stars are destroyed, and that the unbound stars are replenished
from a Maxwellian reservoir,
\begin{equation}
g_{M}(x\!>\! x_{D})\!=\!0\,,\quad g_{M}(x\!<\!0)\!=\! C_{M}\exp[(\sigma_{\star}^{2}/\sigma_{M}^{2})x]\,,\label{e:BCs}
\end{equation}
 where the constant $C_{M}$ is related to $S_{M}$ (the asymptotic number density ratio of star M relative to the reference star; $S_* = 1$ by definition) by $C_{M}=\left(\sigma_{\star}/\sigma_{M}\right)^{3}S_{M}\,$.
They assume violent relaxation boundary conditions such that $C_{M}\!=\! S_{M}\!=\! S_{\star}$; the steady state DFs do not depend strongly on these specific choices of boundary conditions \citep{Bah77, Ale09}.

\begin{figure}[!h]
\begin{center}
\includegraphics[height=85 mm,angle=0 ]{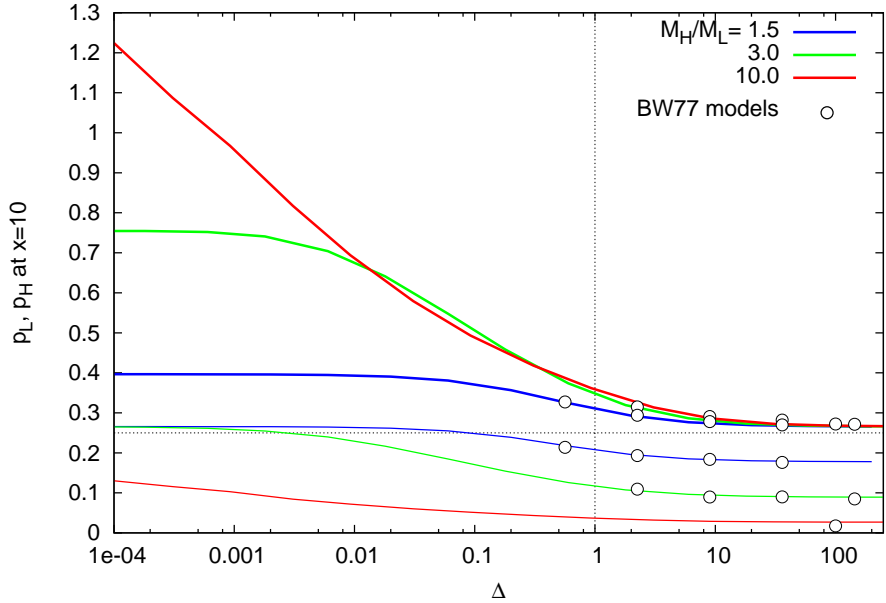}
\caption{The power-law indices $P_{\rm L,H}$ (thin and thick lines respectively), as derived from the logarithmic slopes of $g_{\rm L,H}$ at $x = 10$ as a function of $\Delta$ for different mass ratios. The logarithmic slopes of the DFs as calculated by \citet{Bah77} are over-plotted as open circles. The transition between the weak and strong mass segregation solutions at $\Delta \sim 1$, which is a reflection of the breakdown of the zero-flow assumption as $\Delta \rightarrow 0$, is marked. {\it Figure reproduced from \citet{Ale09} by permission of the AAS.}
\label{f:Delta}}
\end{center}
\end{figure}

These calculations are used to explore steady state solutions for stellar distributions in both the weak and strong mass segregation regime, the nature of which they notate using the relaxational coupling parameter $\Delta$. This parameter describes the competition between the self-coupling of the heavy stars and the light-heavy coupling in terms of the mass and number ratios,
\begin{equation}
\Delta = {N_H M^2_H \over N_L M^2_L } \times {4 \over 3 + M_H/M_L}.
\label{e:Delta}
\end{equation}
We reproduce some results of their simulations in Figure (\ref{f:Delta}) which shows the local logarithmic slopes of the DFs, $p_L$ and $p_H$, at $x=10$ (which corresponds to orbits with a semi major axis of $a = 0.1 $pc for $r_* = 2$ pc in the GC) as a function of $\Delta$. For comparison, they take the same mass ratios as modeled by BW77, $M_H /M_L = 1.5, 3, 10$. 

In the weak segregation limit (the Bahcall-Wolf solution), $\Delta \rightarrow \infty$, which is the zero-flow ($Q_M \rightarrow 0$) limit, the heavy stars dominate the population and relax to the single mass cusp $\alpha_H = 7/4$ ($p_H=1/4$). The light stars heat by scattering against the effectively infinite reservoir of heavy stars and diffuse to lower energies, thereby settling to a flatter cusp with $\alpha_L \rightarrow 3/2$ ($p_L = M_L/4M_H \rightarrow 0$). In the strong segregation limit, $\Delta \rightarrow 0$, the light stars, which dominate the population, behave as a single mass population with $\alpha_L = 7/4$ ($p_L =1/4$). The rare heavy stars sink to the centre by dynamical friction against the effectively infinite reservoir of light stars. For low mass ratios, $M_H /M_L < 4$, where dynamical friction is less efficient, the heavy stars approximately obey the BW77 relation, $p_H= (M_H /M_L) p_L = M_H /4M_L$. For higher mass ratios, $M_H /M_L > 4$, the heavy stars approach the dynamical friction limit, $p_H \rightarrow 5/4$. \\

The present-day mass function of evolved stellar populations (coeval or continuously star forming) with a universal initial mass function, separates into two distinct mass scales, $\sim 1 M_{\odot}$ of main sequence stars, dwarfs and neutron stars, and $\sim 10 M_{\odot}$ of BHs. In this sense such systems can be approximated by a duo-mass system. Furthermore, \citet{Ale09} show that $\Delta < 0.1$ so that this system should be strongly mass segregated. In the next section we calculate the time scale for such a system to form a mass segregated steady state cusp.

\section{Time-dependence of cusp formation}\label{s:time}

\begin{figure}[!h]
\begin{center}
\includegraphics[height=120 mm,angle=270 ]{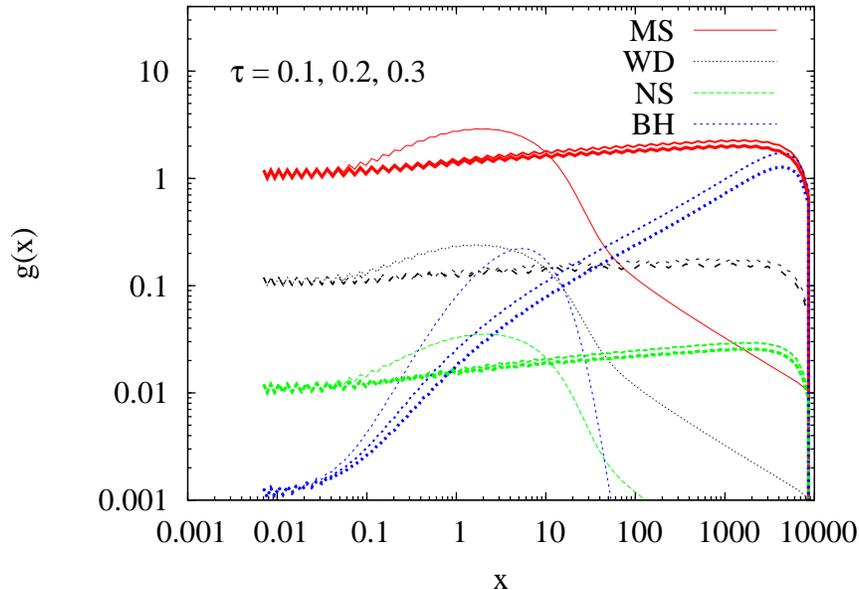}
\caption{Evolution of the distribution function, shown at dimensionless times $\tau=\{0.1, 0.2, 0.3\}$ with increasing thickness for increasing time. Steady state is reached in about $\tau=0.2-0.3$. Initially, all DFs were $g(x)\propto x^{-0.5}$. Interestingly, for short times the distribution of white dwarfs (WD), main sequence stars (MS) and neutron stars (NS) grows steeper than the final distribution which is approximately energy independent. The reason is that it takes time for the stellar black holes (BH) to grow a steep cusp and become the dominant species. \label{f:ss}}
\end{center}
\end{figure}

The amount of time it takes to form a stellar cusp around a MBH in a galactic nucleus is important for many reasons including, but not limited to, event rates of stellar collisions, tidal disruptions of stars and inspirals of compact objects with the emission of gravitational waves. In particular, if this time is greater than a Hubble time, event rates will be lower than previously predicted. To determine this time scale we use Fokker-Planck simulations as described in \S\ref{s:duo} with initial DFs $g_M(x)\propto x^{-0.5}$. We find that steady state is reached in $\tau = 0.2-0.3$; see Figure (\ref{f:ss}).

Our result is in agreement with \citet{Pre10} who find, using both Fokker-Planck and $N$-body simulations, that steady state is reached in $t \sim (0.1 - 0.2) ~ T_{\rm rlx}(r_h)$, where $T_{\rm rlx}(r_h)$ is the value of the relaxation time at the radius of influence of a MBH, and roughly equivalent to $t_{\star}$ in Equation (\ref{e:tstar}). Values of the relaxation time for MBHs with $M_{\bullet} \lesssim 6 \times 10^6 M_{\odot}$ vary in the literature, but are generally constrained to within $0.5 - 3 \times 10^{10} {\rm yr}$. Our results imply that quasi-steady, mass segregated stellar cusps should be common around MBHs in galactic nuclei for this mass regime. Hence, assuming GC parameters that we observe today \citep[see][]{Mer09b}, a cusp should form within 
\begin{eqnarray}
t_{\rm cusp} & = & 0.2 -0.3~ t_{\star} \nonumber\\
 & \approxeq & 0.5 - 6  \times 10^9 {\rm yr} .
   \label{e:tcusp}
 \end{eqnarray} 

\section{Applications}

\subsection{Galactic centre cusp}

Several studies \citep[e.g.][]{Ale99a, Gen03a, Sch07} have shown that indeed a cusp of stars is present in the GC. However, recently it has become clear that even though the {\it young} stars do have a power law profile, there is a dearth of giant stars within $\sim0.1{\rm pc}$ \citep{Buc09, Do09, Bar10}. Since the O/B stars are too young to have relaxed by the two-body relaxation process discussed here, the conclusion is that currently there is no observational support for cusp formation or mass-segregation in the GC. The reason for the absence of a cusp is unclear. One possibility is that physics not considered here depletes the tightly bound orbits of old stars, for example as a result of stellar collisions \citep[e.g.,][]{Dal09}. 

Another option was considered by \citet{Mer09b}, who suggested that relaxational physics {\it does} in fact capture the main mechanism for cusp formation, but that the age of the GC is too short compared to the time scale for a cusp to have formed today. His estimate of the relaxation time is $\gtrsim 2 - 3 \times 10^{10} ~{\rm yr}$, and in his Fokker-Planck simulations, \citet{Mer09b} does not find the formation of a relaxed cusp within that time. This is in contrast to the result we found in \S\ref{s:time}, where a cusp forms in $0.2-0.3 ~ t_{\star}$, similar to the findings by \citet{Pre10}. Thus, even for a relaxation time of $30$ Gyr, the central pc would form a cusp in less than a Hubble time, {\it starting from the current situation}. The reason for this discrepancy in the time scale for reaching a steady state solution is not yet clear, but may be related to the back reaction of the BHs on the light main sequence stars \citep{Pre10}. We note that our findings do not imply that the GC is necessarily relaxed and formed a cusp: it is possible that in the past the density at the radius of influence was lower, that it has taken a Hubble time to contract\footnote{Our treatment of the Fokker-Planck equations does not evolve the stars that are unbound to the MBH, such that we cannot follow the contraction of the cusp. For a study that includes self-consistent mass-segregation and evolution of the ambient cluster of stars, see \citet{Pre10}.} to the current configuration, and that the remaining time to form a cusp is less than a Hubble time \citep{Mer09b}. However, this argument appears to be more attractive when the remaining time for cusp formation exceeds a Hubble time.

\subsection{S-stars}

Relaxational dynamics cannot account for the presence of a cluster of B-stars known as the``S-stars" in the inner $\sim0.03 {\rm pc}$ near the MBH in the GC \citep{Sch02, Ghe05, Eis05, Gil09}, since the time scales are too long for young stars (with ages between $\sim 20 -100$ Myr) to diffuse to such distances. There have been many suggestions of how the S-stars have reached their current orbits. Currently the most promising scenarios seem to be those where a binary star is tidally disrupted by the MBH \citep{Hil88, YuQ03}. Such binaries may originate very far from the MBH, where they are perhaps driven to eccentric orbits by triaxiality or massive perturbers \citep{Per07}. Alternatively, binaries which form in eccentric accretion discs can be pumped to very high eccentricities by a gravitational instability \citep{Mad09}. In either case, a B-star may end up on a tightly bound orbit, but the orbit will be very eccentric ($e\!\sim\!0.99$). Post-capture dynamics must then account for the mapping of the initial, very high eccentricity distribution to the current distribution in the GC.

The mechanism responsible for changing the eccentricities is most likely resonant relaxation \citep[RR;][]{Rau96}, a secular process that very efficiently randomizes the angular momenta of orbits if the precession time is much longer than the orbital time-scale. For the S-stars, the resonant relaxation time is of order
\begin{equation}
T_{\rm RR} \sim 100\, {\rm Myr} {M_{\odot}\over M_{\star}},
\end{equation}
where $M_{\star}$ is the typical mass of stars. For $M_{\star}=10M_{\odot}$, appropriate for mass segregated BHs, this time-scale is short enough to randomize the orbits \citep{Hop06, Lev07}. Post-capture evolution was studied in detail by \citet{Per09} with the use of $N$-body simulations. These simulations do not include general relativistic precession, which is potentially important for the highly eccentric orbits that are considered here, but they clearly show that the distribution evolves on time-scales of the order $T_{\rm RR}$. For $M_{\star} = 1 M_{\odot}$ however, this time scale is too large for the S-star orbits to have evolved from high eccentricities to their currently observed orbits. Only if there are mass-segregated BHs within the S-star radii can RR explain their orbital parameters.

\subsection{Gravitational waves}
As stated in the introduction, most compact remnants that spiral in successfully onto MBHs to become detectable gravitational wave sources originate from distances very close ($\lesssim0.01 {\rm pc}$) to the MBH. It is therefore important to what extent such orbits are populated, and mass-segregation helps to increase this number. In spite of this, even if there is no cusp in the GC, this would not necessarily imply that other, similar galaxies do not have cusps either. If, as we find, the time for cusp formation from the current state is less than a Hubble time, then one would expect that most galaxies with MBHs similar to the GC did in fact form cusps, even if our Galaxy did not. Furthermore, most inspiral sources originate from lower mass MBHs \citep{Hop09, Gai09}; such galaxies have even shorter relaxation times, and are thus more likely to have formed cusps. The predicted number of detectable LISA sources will therefore not depend strongly on the question of whether a cusp is present or not in the GC. Conclusions regarding the absence of a gravitational wave background from fly-bys \citep{Too09} will likewise not be strongly affected.


\begin{thebibliography}
\expandafter\ifx\csname natexlab\endcsname\relax\def\natexlab#1{#1}\fi

\bibitem[{{Alexander}(1999)}]{Ale99a}
{Alexander}, T. 1999, \apj, 527, 835

\bibitem[{{Alexander} \& {Hopman}(2009)}]{Ale09}
{Alexander}, T., \& {Hopman}, C. 2009, \apj, 697, 1861

\bibitem[{{Bahcall} \& {Wolf}(1976)}]{Bah76}
{Bahcall}, J.~N., \& {Wolf}, R.~A. 1976, \apj, 209, 214

\bibitem[{{Bahcall} \& {Wolf}(1977)}]{Bah77}
---. 1977, \apj, 216, 883

\bibitem[{{Bartko} {et~al.}(2010){Bartko}, {Martins}, {Trippe}, {Fritz},
  {Genzel}, {Ott}, {Eisenhauer}, {Gillessen}, {Paumard}, {Alexander},
  {Dodds-Eden}, {Gerhard}, {Levin}, {Mascetti}, {Nayakshin}, {Perets},
  {Perrin}, {Pfuhl}, {Reid}, {Rouan}, {Zilka}, \& {Sternberg}}]{Bar10}
{Bartko}, H., {et~al.} 2010, \apj, 708, 834

\bibitem[{{Baumgardt} {et~al.}(2004){Baumgardt}, {Makino}, \&
  {Ebisuzaki}}]{Bau04a}
{Baumgardt}, H., {Makino}, J., \& {Ebisuzaki}, T. 2004, \apj, 613, 1133

\bibitem[{{Binney} \& {Tremaine}(2008)}]{Bin08}
{Binney}, J., \& {Tremaine}, S. 2008, Galactic Dynamics, Second edition
  (Princeton, NJ: Princeton University Press)

\bibitem[{{Buchholz} {et~al.}(2009){Buchholz}, {Sch{\"o}del}, \&
  {Eckart}}]{Buc09}
{Buchholz}, R.~M., {Sch{\"o}del}, R., \& {Eckart}, A. 2009, \aap, 499, 483

\bibitem[{{Cohn} \& {Kulsrud}(1978)}]{Coh78}
{Cohn}, H., \& {Kulsrud}, R.~M. 1978, \apj, 226, 1087

\bibitem[{{Dale} {et~al.}(2009){Dale}, {Davies}, {Church}, \&
  {Freitag}}]{Dal09}
{Dale}, J.~E., {Davies}, M.~B., {Church}, R.~P., \& {Freitag}, M. 2009, \mnras,
  393, 1016

\bibitem[{{Do} {et~al.}(2009){Do}, {Ghez}, {Morris}, {Lu}, {Matthews}, {Yelda},
  \& {Larkin}}]{Do09}
{Do}, T., {et~al.} 2009, \apj, 703, 1323

\bibitem[{{Eisenhauer} {et~al.}(2005)}]{Eis05}
{Eisenhauer}, F., {et~al.} 2005, \apj, 628, 246

\bibitem[{{Freitag}(2003)}]{Fre03}
{Freitag}, M. 2003, \apjl, 583, L21

\bibitem[{{Freitag} {et~al.}(2006){Freitag}, {Amaro-Seoane}, \&
  {Kalogera}}]{Fre06}
{Freitag}, M., {Amaro-Seoane}, P., \& {Kalogera}, V. 2006, \apj, 649, 91

\bibitem[{{Freitag} \& {Benz}(2002)}]{Fre02}
{Freitag}, M., \& {Benz}, W. 2002, \aap, 394, 345

\bibitem[{{Gair}(2009)}]{Gai09}
{Gair}, J.~R. 2009, Classical and Quantum Gravity, 26, 094034

\bibitem[{{Genzel} {et~al.}(2003)}]{Gen03a}
{Genzel}, R., {et~al.} 2003, \apj, 594, 812

\bibitem[{{Ghez} {et~al.}(2005){Ghez}, {Salim}, {Hornstein}, {Tanner}, {Lu},
  {Morris}, {Becklin}, \& {Duch{\^ e}ne}}]{Ghe05}
{Ghez}, A.~M., {et~al.} 2005, \apj, 620, 744

\bibitem[{{Gillessen} {et~al.}(2009){Gillessen}, {Eisenhauer}, {Trippe},
  {Alexander}, {Genzel}, {Martins}, \& {Ott}}]{Gil09}
{Gillessen}, S., {et~al.}, 2009, \apj, 692, 1075

\bibitem[{{Hills}(1988)}]{Hil88}
{Hills}, J.~G. 1988, \nat, 331, 687

\bibitem[{{Hopman} \& {Alexander}(2005)}]{Hop05}
{Hopman}, C., \& {Alexander}, T. 2005, \apj, 629, 362

\bibitem[{{Hopman} \& {Alexander}(2006{\natexlab{a}})}]{Hop06}
---. 2006{\natexlab{a}}, \apj, 645, 1152

\bibitem[{{Hopman} \& {Alexander}(2006{\natexlab{b}})}]{Hop06b}
---. 2006{\natexlab{b}}, \apjl, 645, L133

\bibitem[{{Hopman} {et~al.}(2007){Hopman}, {Freitag}, \& {Larson}}]{Hop07}
{Hopman}, C., {Freitag}, M., \& {Larson}, S.~L. 2007, \mnras, 378, 129

\bibitem[{{Hopman}(2009)}]{Hop09}
{Hopman}, C. 2009, Classical and Quantum Gravity, 26, 094028

\bibitem[{{Keshet} {et~al.}(2009){Keshet}, {Hopman}, \& {Alexander}}]{Kes09}
{Keshet}, U., {Hopman}, C., \& {Alexander}, T. 2009, \apjl, 698, L64

\bibitem[{{Levin}(2007)}]{Lev07}
{Levin}, Y. 2007, \mnras, 374, 515

\bibitem[{{Madigan} {et~al.}(2009){Madigan}, {Levin}, \& {Hopman}}]{Mad09}
{Madigan}, A.-M., {Levin}, Y., \& {Hopman}, C. 2009, \apjl, 697, L44

\bibitem[{{Merritt}(2009)}]{Mer09b}
{Merritt}, D. 2009, ArXiv e-prints

\bibitem[{{Murphy} {et~al.}(1991){Murphy}, {Cohn}, \& {Durisen}}]{Mur91}
{Murphy}, B.~W., {Cohn}, H.~N., \& {Durisen}, R.~H. 1991, \apj, 370, 60

\bibitem[{{Peebles}(1972)}]{Pee72}
{Peebles}, P.~J.~E. 1972, \apj, 178, 371

\bibitem[{{Perets}(2009)}]{Per09}
{Perets}, H.~B. 2009, \apj, 690, 795

\bibitem[{{Perets} {et~al.}(2007){Perets}, {Hopman}, \& {Alexander}}]{Per07}
{Perets}, H.~B., {Hopman}, C., \& {Alexander}, T. 2007, \apj, 656, 709

\bibitem[{{Preto} \& {Amaro-Seoane}(2010)}]{Pre10}
{Preto}, M., \& {Amaro-Seoane}, P. 2010, \apjl, 708, L42

\bibitem[{{Preto} {et~al.}(2004){Preto}, {Merritt}, \& {Spurzem}}]{Pre04}
{Preto}, M., {Merritt}, D., \& {Spurzem}, R. 2004, \apjl, 613, L109

\bibitem[{{Rauch} \& {Tremaine}(1996)}]{Rau96}
{Rauch}, K.~P., \& {Tremaine}, S. 1996, New Astronomy, 1, 149

\bibitem[{{Rubbo} {et~al.}(2006){Rubbo}, {Holley-Bockelmann}, \&
  {Finn}}]{Rub06}
{Rubbo}, L.~J., {Holley-Bockelmann}, K., \& {Finn}, L.~S. 2006, \apjl, 649, L25

\bibitem[{{Sari} \& {Goldreich}(2006)}]{Sar06}
{Sari}, R., \& {Goldreich}, P. 2006, \apjl, 642, L65

\bibitem[{{Sch{\"o}del} {et~al.}(2007){Sch{\"o}del}, {Eckart}, {Alexander},
  {Merritt}, {Genzel}, {Sternberg}, {Meyer}, {Kul}, {Moultaka}, {Ott}, \&
  {Straubmeier}}]{Sch07}
{Sch{\"o}del}, R., {et~al.} 2007, \aap, 469, 125

\bibitem[{{Sch{\"o}del} {et~al.}(2002)}]{Sch02}
{Sch{\"o}del}, R., {et~al.} 2002, \nat, 419, 694

\bibitem[{{Shapiro} \& {Marchant}(1978)}]{Sha78}
{Shapiro}, S.~L., \& {Marchant}, A.~B. 1978, \apj, 225, 603

\bibitem[{{Sigurdsson} \& {Rees}(1997)}]{Sig97b}
{Sigurdsson}, S., \& {Rees}, M.~J. 1997, \mnras, 284, 318


\bibitem[{{Toonen} {et~al.}(2009){Toonen}, {Hopman}, \& {Freitag}}]{Too09}
{Toonen}, S., {Hopman}, C., \& {Freitag}, M. 2009, \mnras, 398, 1228

\bibitem[{{Yu} \& {Tremaine}(2003)}]{YuQ03}
{Yu}, Q., \& {Tremaine}, S. 2003, \apj, 599, 1129

\end{thebibliography}
\end{document}